\documentclass{article}
\usepackage{spconf, amsmath, graphicx, hyperref}
\usepackage{booktabs, multirow, makecell, xcolor, pifont}

\newcommand{\cmark}{\ding{51}}
\newcommand{\xmark}{\ding{55}}
\title{RLBR: Reinforcement Learning with Biasing Rewards for Contextual Speech Large
Language Models}
\name{Bo Ren, Ruchao Fan, Yelong Shen, Weizhu Chen, Jinyu Li}
\address{Microsoft Core AI, USA}
\begin{document}
    \ninept
    \maketitle

    \begin{abstract}
        Speech large language models (LLMs) have driven significant progress in end-to-end
        speech understanding and recognition, yet they continue to struggle with
        accurately recognizing rare words and domain-specific terminology. This paper
        presents a novel fine-tuning method, \textbf{Reinforcement Learning with
        Biasing Rewards (RLBR)}, which employs a specialized biasing words
        preferred reward to explicitly emphasize biasing words in the reward
        calculation. In addition, we introduce \textbf{reference-aware
        mechanisms} that extend the reinforcement learning algorithm with reference
        transcription to strengthen the potential trajectory exploration space.
        Experiments on the LibriSpeech corpus across various biasing list sizes
        demonstrate that RLBR delivers substantial performance improvements over
        strong supervised fine-tuning (SFT) baseline and consistently
        outperforms several recently published methods. The proposed approach achieves
        excellent performance on the LibriSpeech \emph{test-clean}/\emph{test-other}
        sets, reaching Biasing Word Error Rates (BWERs) of \textbf{0.59\%}/\textbf{2.11\%},
        \textbf{1.09\%}/\textbf{3.24\%} and \textbf{1.36\%}/\textbf{4.04\%} for biasing
        list sizes of 100, 500, and 1000, respectively, without compromising the
        overall WERs.
    \end{abstract}
    \section{Introduction}
    Contextual biasing plays a critical role in Automatic Speech Recognition (ASR)
    by enabling accurate transcription of rare words, named entities, and domain-specific
    terms \cite{li2022recent}. Conventional ASR models tend to favor frequent vocabulary,
    often misrecognizing or omitting important context-dependent words. To address
    this, effective contextual biasing strategies are needed to dynamically
    adapt model predictions using external information or user-provided context,
    ensuring reliable transcription in practical applications.

    The emergence of speech LLMs has substantially advanced general ASR performance~\cite{10389705,audiopalm2023,bai2024seed,abouelenin2025phi,fan2025alignformer}.
    Nevertheless, contextual biasing for rare or domain-specific terms in speech
    LLMs remains insufficiently addressed~\cite{lakomkin2024end,yang2024ctc}.
    Conventional methods, such as shallow fusion~\cite{williams2018contextual,zhao2019shallow},
    deep biasing~\cite{le2021deep,le2021contextualized,pundak2018deep,contextual_adapters_2022},
    dynamic vocabulary injection~\cite{sudo2024contextualized}, contextual
    spelling correction \cite{wang2021a}, and prompt-based contextual ASR~\cite{ren2025lightweight},
    typically require specialized decoding or architectural changes that are not
    readily compatible with LLMs, complicating adaptation and deployment.

    Some recent works have explored fine-tuning speech LLMs to improve the model's
    ability of leveraging external contextual information by embedding contextual
    information into the input prompt~\cite{lakomkin2024end,yang2024ctc,nozawa2024enhancing}.
    For example,~\cite{lakomkin2024end} shows that metadata such as video titles
    or descriptions can be injected into prompts for Contextualized LLM-based ASR.
    Authors of~\cite{yang2024ctc} introduces an efficient filtering method using
    a CTC decoder to filter irrelevant biasing words, reducing computational load
    but serving only as a preprocessing step. These initial studies highlight
    the potential of fine-tuning speech LLMs for contextual biasing, but their performance
    gains remain limited due to indirect optimization via auxiliary objectives,
    either maximizing the likelihood of next token predictions or minimizing
    errors in the input biasing list by filtering irrelevant words. Such indirect
    optimization of target metrics often leads to sub-optimal performance. In
    contrast of the proxy objective optimization, Reinforcement Learning (RL) offers
    an efficient way to directly optimize the models towards the desired target metrics.

    In the domain of text-based LLMs, RL has proven highly effective for
    aligning model outputs with human preferences across a wide range of tasks~\cite{ouyang2022training,dubey2024llama,guo2025deepseek}.
    Proximal Policy Optimization (PPO)~\cite{schulman2017proximal} is commonly used
    to optimize LLMs based on human feedback via proxy reward models. More
    recently, simplified variants of PPO, such as Direct Preference Optimization
    (DPO)~\cite{rafailov2023direct} and Group Relative Policy Optimization (GRPO)~\cite{shao2024deepseekmath},
    have been developed to further streamline and enhance the training process.
    In the context of speech LLMs, RL has been increasingly applied to tasks including
    audio understanding, speech generalization, audio question answering, and
    speech recognition~\cite{bai2024seed,chu2024qwen2,lin2024align,li2025reinforcement}.
    However, these efforts have primarily targeted specific speech tasks. To date,
    there has been no reported attempt to leverage RL specifically for improving
    contextual biasing in speech LLMs.

    In this work, we introduce Reinforcement Learning with Biasing Rewards (RLBR),
    an RL-based fine-tuning method designed to enhance the contextual biasing
    capabilities of speech LLMs. RLBR improves the recognition of biasing words by
    incorporating a specialized reward function that assigns prioritized weight to
    these words within the speech signal. In addition, it extends the RL algorithm
    with reference-aware mechanisms that integrate reference transcriptions into
    the trajectory group, thereby expanding the exploration space during RL
    training. Building upon a high-performing SFT seed model, RLBR further
    improves the model's ability to utilize contextual information for rare term
    recognition. This methodology enables robust and effective contextual adaptation
    without requiring any changes to the model architecture or decoding process,
    thereby opening broad potential for application to other speech tasks such
    as contextual speech translation. The main contributions of this work are summarized
    as follows:
    \begin{itemize}
        \item To the best of our knowledge, this is the first work to apply RL
            to enhance contextual biasing in speech LLMs.

        \item We propose the RLBR fine-tuning method, which prioritizes the
            accurate recognition of biasing words and leads to substantial
            improvements in contextual biasing performance.

        \item We further enhance RLBR by introducing reference-aware mechanisms that
            strengthen the trajectory exploration space during training, leading
            to consistent performance gains.
    \end{itemize}

    \section{Method}
    In this section, we describe our proposed RLBR method for contextual biasing
    in speech LLMs. We begin with a concise overview of the backbone GRPO algorithm
    utilized in this work. Next, we detail the application of GRPO to contextual
    biasing, followed by an explanation of the RLBR method and its extension
    through reference-aware mechanisms.
    \newcommand{\ED}{\mathcal{ED}}

    \begin{figure}[t!]
        \centering
        \includegraphics[width=\linewidth]{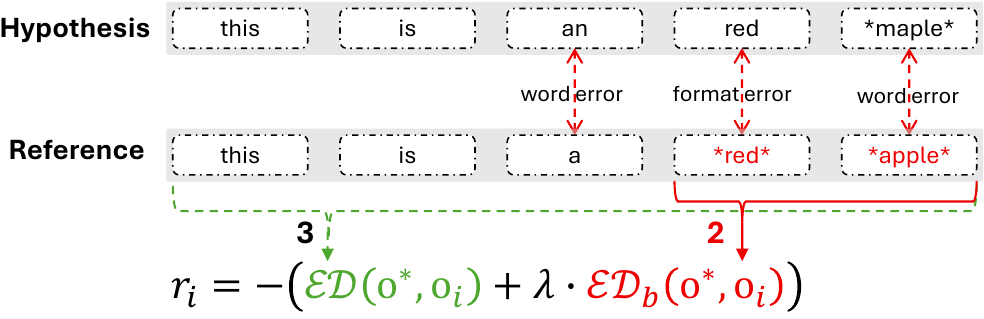}
        \caption{Illustration of the biasing word preferred reward rule. Word-level
        edit distance is shown: $\textcolor{green}{\ED(o^{*}, o_{i})}=3$ for all
        words, and $\textcolor{red}{\ED_{b}(o^{*}, o_{i})}=2$ for biasing words only.}
        \label{fig:reward}
    \end{figure}
    \subsection{Group Relative Policy Optimization}
    \label{sec:grpo}

    The Group Relative Policy Optimization (GRPO) algorithm~\cite{shao2024deepseekmath}
    is a simplified variant of PPO that achieves comparable performance while streamlining
    the training process. GRPO eliminates the need of the value model and allow
    rewards to be computed directly using rule-based methods, which simplifies implementation
    and enhances training stability. As a result, GRPO can be efficiently applied
    to a wide range of LLM tasks, providing a streamlined approach for aligning
    model outputs with user preferences.

    The model parameters are updated by maximizing the GRPO objective ~\cite{shao2024deepseekmath}:
    \newcommand{\piDelta}{\frac{\pi_{\theta}\big(o_{i,t}\mid q, o_{i,<t}\big)}{\pi_{\theta_{\mathrm{old}}}\big(o_{i,t}\mid
    q, o_{i,<t}\big)}}
    \begin{equation}
        \label{eq:grpo}
        \begin{aligned}
            \mathcal{J}_{\mathrm{GRPO}}(\theta) & = \frac{1}{G}\sum_{i=1}^{G}\frac{1}{|o_i|}\sum_{t=1}^{|o_i|}\Big\{ \min \Big[ \piDelta A_{i,t},\; \\
                                                & \quad \operatorname{clip}\big( \piDelta,\; 1-\epsilon,\; 1+\epsilon \big) A_{i,t}\Big]            \\
                                                & \quad - \beta D_{\mathrm{KL}}\big[ \pi_{\theta}\,\|\, \pi_{\mathrm{ref}}\big] \Big\}
        \end{aligned}
    \end{equation}
    where $G$ denotes the number of generated trajectories sampled from the old
    model policy $\pi_{\theta_{\mathrm{old}}}$ and the same question prompt $q$.
    $\pi_{\theta_{\mathrm{old}}}$ is the previous version of the current model policy
    $\pi_{\theta}$ during training iterations. $o_{i}$ represents the $i$th generated
    trajectory, and $A_{i,t}$ is the advantage defined as
    \begin{equation}
        A_{i,t}= \frac{r_{i}- \operatorname{mean}(R)}{\operatorname{std}(R)}
        \centering
        \label{eq:rel_advantage}
    \end{equation}
    where $r_{i}$ denotes the reward assigned to the trajectory $o_{i}$, and $R =
    \{r_{0}, r_{1}, \cdots, r_{G}\}$ represents the set of rewards for the grouped
    trajectories corresponding to the same question prompt $q$. The mean and
    standard deviation are calculated across this group to normalize the rewards.
    The parameter $\epsilon$ is used to clip the policy ratio, which helps
    stabilize the training process. The term $\beta D_{\mathrm{KL}}[ \pi_{\theta}
    \,\| \, \pi_{\mathrm{ref}}]$ serves as a regularization, controlling the
    divergence between the current policy model $\pi_{\theta}$ and the reference
    model $\pi_{\mathrm{ref}}$. In our experiments, we set $\beta = 0$ based on
    preliminary findings that this regularization has minimal impact on
    performance. The value of $\epsilon$ is set to 0.28, following the
    recommendations in \cite{yu2025dapo}.

    \subsection{GRPO for Contextual biasing}
    \label{sec:grpo_cb}

    RL techniques provide an effective means to directly optimize models with respect
    to human preferences. For contextual biasing tasks, RL is a natural fit;
    however, selecting the appropriate RL algorithm is essential. A key factor
    in successful policy optimization is the design of an efficient reward
    function. In natural language tasks, the quality measurement of model
    response is challenging due to the inherent flexibility of languages. In contrast,
    contextual biasing tasks benefit from human-annotated transcriptions as
    expected responses, which are deterministic, clearly defined, and can be reliably
    assessed using rule-based metrics such as edit-distance. This property makes
    GRPO particularly well-suited for contextual biasing, as it leverages rule-based
    rewards to enable efficient and stable policy optimization.

    GRPO can be directly applied to speech LLMs for contextual biasing by utilizing
    a standard edit-distance based reward function. Specifically, for each input
    consisting of a text prompt and speech signal, the Speech LLM generates a
    set of $G$ hypotheses via categorical sampling, denoted as $\mathrm{O}= \{o_{0}
    , o_{1}, \cdots, o_{G}\}$. For each hypothesis $o_{i}$, the reward $r_{i}$ is
    computed as follows:
    \begin{equation}
        r_{i}= - \ED(o^{*}, o_{i})
        \centering
        \label{eq:standard_reward}
    \end{equation}
    Here, $\ED(o^{*}, o_{i})$ represents the edit-distance between the human reference
    transcription $o^{*}$ and the hypothesis $o_{i}$. The model parameters are
    updated by maximizing the GRPO objective in Equation~\ref{eq:grpo} using the
    set of computed rewards $R$.

    \begin{figure}[!t]
        \centering
        \includegraphics[width=\linewidth]{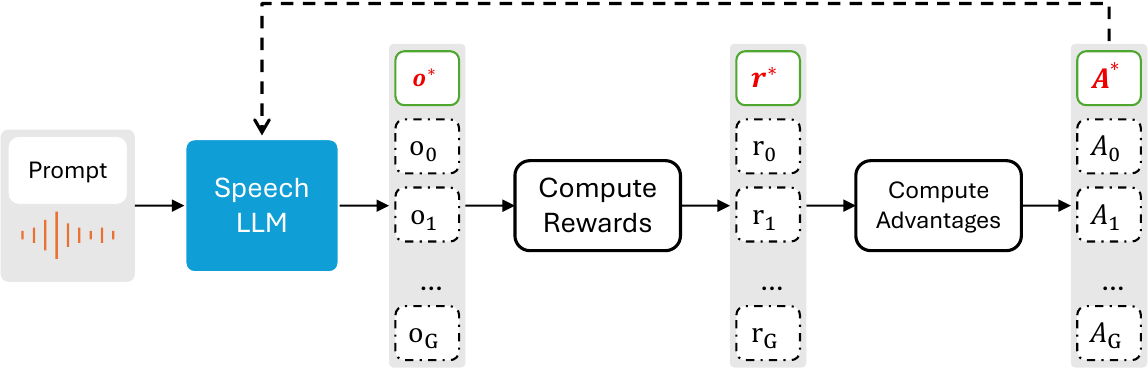}
        \caption{Reference-aware GRPO for Speech LLMs.}
        \label{fig:grpo}
    \end{figure}

    \begin{table*}
        [t!]
        \centering
        \caption{BWER (WER/UWER) (\%) results of different methods under various
        biasing list sizes ($N$). Bold values indicate the best \textbf{BWER}
        per column.}
        \label{tab:main_result}
        \vspace{0.1in}
        \setlength{\tabcolsep}{6pt}
        \renewcommand{\arraystretch}{1.15}
        \begin{tabular}{l|cc|cc|cc|cc}
            \toprule \multirow{2}{*}{Method}                                         & \multicolumn{2}{c|}{$N=0$}            & \multicolumn{2}{c|}{$N=100$}           & \multicolumn{2}{c|}{$N=500$}          & \multicolumn{2}{c}{$N=1000$}           \\
            \cmidrule(lr){2-3}\cmidrule(lr){4-5}\cmidrule(lr){6-7}\cmidrule(lr){8-9} & test-clean                            & test-other                             & test-clean                            & test-other                            & test-clean                            & test-other                            & test-clean                            & test-other                            \\
            \midrule \makecell[l]{Dynamic\\Vocabulary~\cite{sudo2024contextualized}} & \makecell{13.80\\(3.16/1.90)}         & \makecell{27.50\\(6.95/4.60)}          & \makecell{2.80\\(1.80/1.70)}          & \makecell{7.10\\(4.63/4.30)}          & \makecell{3.10\\(1.92/1.80)}          & \makecell{7.90\\(4.81/4.50)}          & \makecell{3.30\\(2.01/1.90)}          & \makecell{8.50\\(4.97/4.60)}          \\
            \midrule \makecell[l]{CTC-Assisted\\LLM~\cite{yang2024ctc}}              & \makecell{9.33\\(1.96/1.11)}          & \makecell{20.02\\(4.18/2.49)}          & \makecell{3.67\\(1.27/1.00)}          & \makecell{8.02\\(2.72/2.16)}          & \makecell{3.92\\(1.33/1.03)}          & \makecell{9.04\\(3.04/2.40)}          & \makecell{4.16\\(1.33/1.00)}          & \makecell{9.33\\(2.99/2.31)}          \\
            \midrule\midrule Phi-4-Multimodal~\cite{abouelenin2025phi}               & \makecell{\textbf{7.44}\\(1.67/0.94)} & \makecell{\textbf{16.97}\\(3.87/2.32)} & \makecell{5.89\\(1.48/0.92)}          & \makecell{12.90\\(3.43/2.31)}         & \makecell{37.87\\(116/126)}           & \makecell{52.85\\(153/165)}           & \makecell{86.44\\(201/215)}           & \makecell{91.91\\(229/245)}           \\
            \cmidrule(lr){2-9} \quad+SFT                                             & \makecell{7.49\\(1.72/0.99)}          & \makecell{17.20\\(3.96/2.40)}          & \makecell{1.06\\(0.95/0.93)}          & \makecell{2.94\\(2.35/2.28)}          & \makecell{1.70\\(1.06/0.98)}          & \makecell{5.17\\(2.71/2.42)}          & \makecell{2.38\\(1.14/0.99)}          & \makecell{6.41\\(2.88/2.47)}          \\
            \cmidrule(lr){2-9} \qquad++RLBR                                          & \makecell{7.82\\(1.72/0.95)}          & \makecell{17.37\\(4.00/2.42)}          & \makecell{\textbf{0.59}\\(0.82/0.85)} & \makecell{\textbf{2.11}\\(2.25/2.26)} & \makecell{\textbf{1.09}\\(0.96/0.95)} & \makecell{\textbf{3.24}\\(2.52/2.44)} & \makecell{\textbf{1.36}\\(1.07/1.03)} & \makecell{\textbf{4.04}\\(2.65/2.49)} \\
            \bottomrule
        \end{tabular}
    \end{table*}

    \subsection{Reinforcement Learning with Biasing Rewards}
    Although standard edit-distance based rewards in Equation~\ref{eq:standard_reward}
    are effective for measuring overall hypothesis quality, they treat all edit
    errors equally without specific emphasis on the highly concerned biasing
    words in contextual biasing task, make standard edit-distance rewards less suitable.
    To address this limitation, we propose the Reinforcement Learning with Biasing
    Rewards (RLBR) method, which incorporates a specialized biasing word
    preferred reward rule that explicitly increases the contribution of biasing
    words in the reward calculation. In this approach, edit-distances for both
    all the words and biasing words are combined using a weighted scheme, as illustrated
    in Figure~\ref{fig:reward}. For each hypothesis $o_{i}$, the reward is computed
    as:
    \begin{equation}
        r_{i}= - (\ED(o^{*}, o_{i})+ \lambda \cdot \ED_{b}(o^{*}, o_{i}))
        \centering
        \label{eq:bias_reward}
    \end{equation}
    where $\ED_{b}(o^{*}, o_{i})$ computes the edit-distance specifically for
    biasing words in the reference transcription against their best-matching
    spans in hypothesis. The weighting factor $\lambda$ controls the relative importance
    of biasing words in the reward calculation. Increasing $\lambda$ places
    greater emphasis on correctly recognizing biasing terms, guiding the model to
    focus more on these words during training.

    The edit-distance for reward calculation can be evaluated at either word or character
    level, providing flexible control over feedback granularity. Our ablation study
    shows that character-level feedback consistently leads to improved recognition
    of biasing words, likely because it captures the minor differences between
    hypothesis and transcription more effectively.

    As illustrated in Figure~\ref{fig:reward}, biasing words are clearly marked with
    special tags (e.g., $*$) to distinguish them from general vocabulary. This
    explicit formatting enables the reward function to reliably identify and
    evaluate the model's handling of biasing words, ensuring that these targeted
    biasing words receive appropriate emphasis during training. Our ablation
    studies confirm that this biasing word formatting consistently yields better
    performance than using raw text.

    \subsection{Reference-aware GRPO}
    As discussed, GRPO relies on grouped hypotheses $\mathrm{O}$ sampled from the
    seed model to compute the advantage for policy optimization. These
    hypotheses are essential for providing a diverse exploration space during
    training. However, the quality of these hypotheses may be limited by the
    seed model, especially in challenging cases where the model fails to generate
    any correct hypotheses. To address this limitation, we propose a reference-aware
    mechanism that includes the reference transcription $o^{*}$ as an additional
    hypothesis within the hypotheses group $\mathrm{O}^{+}=\{\mathrm{O}, o^{*}\}$,
    as illustrated in Figure~\ref{fig:grpo}. The reward set $R$ is expanded to incorporate
    the reward $r^{*}$ for the reference transcription $o^{*}$, resulting in
    $R^{+}= \{R, r^{*}\}$. The relative advantage $A_{i,t}$ is then calculated based
    on the extended set $R^{+}$ for each hypothesis $o_{i}\in \mathrm{O}^{+}$.
    The ablation studies demonstrate that the reference-aware mechanism
    consistently improves performance across various biasing list sizes.

    \section{Experimental Setup}
    \subsection{Datasets}
    Our experiments are conducted using the LibriSpeech corpus, following established
    protocols from previous studies~\cite{yang2024ctc,le2021contextualized,sudo2024contextualized}.
    The speech LLM-based ASR system is fine-tuned on the full 960-hour
    LibriSpeech training set and evaluated on the standard test-clean and test-other
    splits. For evaluation, artificial biasing lists are constructed for each
    test sample as described in~\cite{le2021contextualized}. Each biasing list contains:
    (i) rare words present in the reference transcription, and (ii) distractor words
    randomly selected from the pool of rare words in the training set. The
    number of distractors $N$ is systematically varied among
    $\{100, 500, 1000\}$ to enable a thorough analysis across different list
    sizes. To comprehensively assess performance, we report three standard
    metrics: Word Error Rate (WER), which measures overall transcription
    accuracy; Biasing Word Error Rate (BWER), which quantifies errors
    specifically on biasing words; and Unbiased Word Error Rate (UWER), which
    evaluates errors on non-biasing (general) words to reflect performance on
    standard vocabulary.

    \subsection{Contextual data augmentation}
    Contextual biasing aims to improve the recognition of words from a specified
    biasing list; however, suitable training data is often limited. To address this
    limitation, we adopt a contextual data augmentation strategy as follows: For
    each audio-text pair, we construct an associated biasing list by sampling
    words from two categories:
    \begin{itemize}
        \item \textbf{Positive words:} Words present in the reference
            transcription, which the model should prioritize and recognize accurately.

        \item \textbf{Negative words:} Distractor words randomly selected from
            the global training vocabulary, intended to help the model ignore irrelevant
            terms.
    \end{itemize}
    The input prompt for the speech LLM is then formatted as: \textbf{``Transcribe
    the audio clip into text with extra attention to the following words: [biasing
    list]''}. Notably, the biasing words in the prompt are also explicitly marked
    with special tags (e.g., $*$) to distinguish them from general vocabulary.
    \begin{table}[!t]
        \centering
        \caption{WER/BWER (\%) results for different biasing weights ($\lambda$).
        The best \textbf{BWER} is highlighted in bold. Word-level and reference-unaware
        reward settings with biasing format are used in the table.}
        \vspace{0.1in}
        \label{tab:bias_weight}
        \begin{tabular}{c|cc|cc}
            \toprule \multirow{2}{*}{$\lambda$}  & \multicolumn{2}{c|}{N = 100}  & \multicolumn{2}{c}{N = 500}    \\
            \cmidrule(lr){2-3}\cmidrule(lr){4-5} & clean-test                    & other-test                    & clean-test                    & other-test                    \\
            \midrule 0                           & \makecell{0.87/\textbf{0.70}} & \makecell{2.67/2.47}          & \makecell{1.06/1.34}          & \makecell{2.85/4.06}          \\
            \midrule 1                           & \makecell{0.87/0.72}          & \makecell{2.66/2.54}          & \makecell{1.05/1.36}          & \makecell{2.80/4.09}          \\
            \midrule 3                           & \makecell{0.88/0.72}          & \makecell{2.37/2.41}          & \makecell{1.08/1.34}          & \makecell{2.80/3.97}          \\
            \midrule 5                           & \makecell{0.87/\textbf{0.70}} & \makecell{2.39/\textbf{2.25}} & \makecell{1.04/\textbf{1.22}} & \makecell{2.82/\textbf{3.74}} \\
            \bottomrule
        \end{tabular}
    \end{table}

    \subsection{Implementation Details}
    In our experiments, we employ Phi-4-Multimodal~\cite{abouelenin2025phi} as
    the backbone speech LLM. This architecture integrates a 3.8B parameter
    language model with a 460M parameter audio encoder. Input speech signals are
    converted into 80 dimensional log Mel filter bank features at a 10 ms frame rate,
    with a maximum duration of 40 seconds.

    As highlighted in \cite{chu2025sft}, SFT is crucial for establishing the core
    capabilities of large language models, while RL further enhances their
    generalization. Prior to applying our RLBR method, we first conduct SFT on
    the Phi-4-Multimodal model using context-augmented data to establish a
    robust seed model that effectively leverages contextual information.

    During both SFT and RLBR training, we employ Low Rank Adaptation (LoRA)~\cite{hu2022lora}
    with a rank size of 320, applied to the attention and feed-forward layers, resulting
    in approximately 460M trainable parameters. The model is updated using the
    complete 960-hour training set with the AdamW optimizer, distributed across 8
    NVIDIA A100 GPUs for both SFT and RLBR stages. We utilize a cosine decay
    learning rate schedule, with a peak learning rate of $1 \times 10^{-5}$ for SFT
    and $5 \times 10^{-6}$ for RLBR fine-tuning.

    During RLBR fine-tuning, we generate 8 hypotheses for each sample using
    categorical sampling with a temperature of 1.2 to encourage hypotheses
    diversity. For optimal performance of RLBR, as shown in Table~\ref{tab:main_result},
    we employ the reference-aware GRPO mechanism together with the biasing word preferred
    reward. Edit-distance for both overall transcription and biasing words is
    computed at the character level where biasing words are formatted with
    special tags. The biasing weight $\lambda$ in Equation~\ref{eq:bias_reward} is
    set to 5, based on ablation studies summarized in Table~\ref{tab:bias_weight}.

    \section{Results and Discussion}
    \subsection{Main Results}
    As shown in Table~\ref{tab:main_result}, we benchmark our proposed method against
    several strong baselines, including Dynamic Vocabulary~\cite{sudo2024contextualized}
    and CTC-Assisted LLM~\cite{yang2024ctc}, which are built on different audio encoders
    and LLM backbones than those used in our method. Across all biasing list
    sizes, our RLBR method consistently delivers substantial improvements across
    all metrics, demonstrating its effectiveness in enhancing contextual biasing
    for speech LLMs.

    The backbone Phi-4-Multimodal model demonstrates strong performance with
    short biasing lists ($N=100$); however, its effectiveness declines as the
    biasing list size increases due to lack of contextual training data, especially
    long context. Incorporating SFT with augmented contextual data enables the
    model to utilize the biasing list more effectively, resulting in significant
    improvements across various biasing list sizes. This underscores the
    critical role of SFT in establishing the model's foundational ability to
    leverage contextual information.

    Compared to the strong SFT baseline, the proposed RLBR method further improves
    model performance by reducing overall WER and, notably, achieving substantial
    relative reductions in BWER (28.2\%--44.3\%) across all biasing list sizes ($N
    > 0$). Importantly, these gains are realized without negatively affecting
    UWER, demonstrating that RLBR fine-tuning enables the model to more accurately
    recognize targeted terms through effective utilization of the biasing list.
    \begin{table}[!t]
        \centering
        \caption{WER/BWER (\%) results with different RLBR strategies on
        LibriSpeech \emph{test-clean} split with biasing weight $\lambda=1$.}
        \label{tab:reward_design}
        \vspace{0.1in}
        \begin{tabular}{l|c|c|c|c|c}
            \toprule Row & \makecell{Edit \\ Level} & \makecell{Reference\\Aware} & \makecell{Biasing\\Format} & N = 100              & N = 500              \\
            \midrule A   & word                     & \xmark                      & \cmark                     & \makecell{0.87/0.72} & \makecell{1.05/1.36} \\
            \midrule B   & word                     & \xmark                      & \xmark                     & \makecell{0.91/1.38} & \makecell{1.02/1.91} \\
            \midrule C   & char                     & \xmark                      & \cmark                     & \makecell{0.87/0.69} & \makecell{1.04/1.33} \\
            \midrule D   & word                     & \cmark                      & \cmark                     & \makecell{0.87/0.69} & \makecell{1.06/1.22} \\
            \bottomrule
        \end{tabular}
    \end{table}
    \subsection{Ablation Study on RLBR}
    \subsubsection{Biasing Weight in Reward}
    We investigated the impact of adjusting the biasing weight ($\lambda$) in
    the reward function, as summarized in Table~\ref{tab:bias_weight}. The
    results indicate that increasing $\lambda$ consistently leads to lower BWER,
    with optimal performance observed at $\lambda=5$. These findings confirm
    that assigning greater importance to biasing words in the reward calculation
    effectively encourages the model to focus on these terms, thereby enhancing contextual
    biasing accuracy.

    \subsubsection{Biasing Words Formatting Impact}
    Table~\ref{tab:reward_design}, rows A and B, demonstrate the effect of explicitly
    biasing words formatting. As shown in Row B, where biasing word formatting is
    omitted, the model experiences a notable increase in BWER compared to Row A.
    This result highlights the importance of clearly identifying biasing terms within
    the reward function, as it enables the model to more effectively focus on
    these critical words.

    \subsubsection{Edit Distance Granularity Choice}
    Rows A and C in Table~\ref{tab:reward_design} illustrate the impact of edit-distance
    granularity in the reward function. Utilizing character-level edit distance consistently
    yields lower BWER for biasing lists of $N \in \{100, 500\}$, as character-level
    rewards provide more detailed and informative feedback by emphasizing
    minimal edits.

    \subsubsection{Reference-aware GRPO}
    The comparison between rows A and D in Table~\ref{tab:reward_design}
    evaluates the impact of including reference transcriptions in the GRPO
    hypothesis group. Adding the reference transcription strengthens the exploration
    space, resulting in improved BWER performance.

    \section{Conclusion}
    In this work, we present RLBR, a novel fine-tuning approach for contextual biasing in speech large language models. By explicitly prioritizing biasing words in the reward function, RLBR substantially improves the recognition of rare and domain specific terms, and be validated on LibriSpeech with various biasing lists. Although not detailed here, RLBR exhibits similar performance gains on internal, realistic domain-specific datasets, further confirming the robustness and effectiveness of the proposed approach.

    \bibliographystyle{IEEEbib}
    \bibliography{refs}
\end{document}